\documentclass[preprintnumbers,amsmath,amssymb,floatfix,10pt,prd,superscriptaddress,nofootinbib]{revtex4}
\usepackage{graphicx}
\usepackage{epsfig}
\usepackage{bm}
\usepackage{amsfonts}
\usepackage{subfigure}
\usepackage{hyperref}

\begin{document}

\title{Effect of  anisotropy on    generalized  Chaplygin gas   scalar  field    and   its interaction with other dark energy models}

\author{V. Fayaz}
\email{fayaz_vahid@yahoo.com}
\affiliation{Department Of Physics, Hamedan Branch, Islamic Azad University, Hamedan, Iran}
\author{H. Hossienkhani}
\email{hossienhossienkhani@yahoo.com}
\affiliation{Department Of Physics, Hamedan Branch, Islamic Azad University, Hamedan, Iran}

\date{\today}
\begin{abstract}
\vspace*{1.5cm} \centerline{\bf Abstract} \vspace*{.5cm}

\end{abstract}


\maketitle

In this work,    we establish a correspondence between the interacting  holographic, new agegraphic dark energy   and  generalized Chaplygin gas  model in  Bianchi type I   universe. In continue, we reconstruct the potential of the scalar field which describes the generalized Chaplygin cosmology. Cosmological solutions are obtained
when the kinetic energy of the phantom field is     order of the anisotropy and dominates over the
potential energy of the field.    We investigate observational constraints on the generalized Chaplygin gas,   holographic and new agegraphic dark energy models as the unification of dark matter and dark energy, by using the latest observational data. To do this we focus on observational determinations of the expansion history $H(z)$. It is shown that the HDE model  is better than the NADE and  generalized  Chaplygin gas    models in  an anisotropic universe. Then, we calculate the
evolution of density perturbations in the linear regime for   three models of dark energy   and compare
the results $\Lambda$CDM model.  Finally, the analysis shows that the increase in anisotropy leads to more correspondence between the dark energy scalar field model and observational data.

{\bf{keywords}:}{ Anisotropic universe,  Holographic dark energy,  New agegraphic dark energy,  Interacting dark energy,   Generalized Chaplygin gas.}


\section{Introduction}
A series of astronomical observations over the past decade indicate that our universe   confirms a state of accelerated expansion \cite{1,2}. Present observational cosmology has provided enough evidence in favour of the accelerated expansion of the universe \cite{3,4,5,6,8}. There exists some unknown energy, which is called dark energy (DE), to realize the  accelerated expansion. A cosmological constant ($\Lambda$) has effective equal pressure
to minus its energy density (equation of state $\omega_\Lambda=-1$) consistent with preliminary measurements, but
 in supersymmetric theories the most natural scale for $\Lambda$ is at least as large as 100 GeV.  So far it  also so-called $\Lambda$CDM,  which provides an excellent fit to a wide range of astronomical data.   As regards, the $\Lambda$CDM model confronts  problems, which are namely ``fine-tuning" and ``cosmic coincidence"  \cite{91,92}. Other the simplest extension of $\Lambda$ is the DE  with a constant $w$,  which is the corresponding cosmological model  so-called  that $wC$DM model \cite{93,94}.
 Recent reviews \cite{10,11,12,13} are useful for a brief knowledge of DE models.
 In recent years, the holographic DE (HDE) has been studied as a possible candidate for DE.  It is  commonly believed that the holographic principle  \cite{14,15,16} is just a fundamental principle of quantum gravity too.  Holographic principle is illuminated by investigations of the quantum property of black holes. In this sense, the number of freedom's degrees  of a physical system should be finite and  scale with its bounding area rather than with its volume. It should be constrained by an infrared cut-off \cite{17}. According to \cite{17} the energy contained in a region of size $L$ must not exceed the mass of a black hole of the same size, which means, in terms of energy density, $\rho_\Lambda\leq L^{-2}$.  Based on this idea,   \cite{18} proposed the HDE model, where the infrared cutoff is taken to be the size of event horizon for DE.  More details about the HDE was studied by many authors \cite{19,20,21,22,23,24,25}.\\
Another proposal to explore the nature of DE within the framework of quantum gravity is  the agegraphic DE (ADE) \cite{26}. This model takes into account the Heisenberg uncertainty relation of quantum mechanics together with the gravitational effect in general relativity. The ADE model considers spacetime and matter field fluctuations responsible for DE. However, the ADE model might contain an inconsistency \cite{27}. So to overcome this problem, which after
the   ADE model, the authors \cite{28} proposed an alternative model of  DE, is namely  the ``new agegraphic DE" (NADE).  The NADE models have been studied in plentiful detail by \cite{29,30}.\\
It was  purposed the use of some perfect fluid with an   equation of state and called it as Chaplygin gas (CG)  \cite{31}. The CG is one of the candidate of DE models to explain the accelerated expansion of the universe. The striking features of CG DE is that it can be assumed as a possible unification of DM and DE. The CG plays a duplex role at different epoch of the history of the universe: it can be as a dust-like matter in the early time (i.e. for small scale factor $a$), and as a cosmological constant at the late time.  Bertolami \textit{et al}. \cite{32} have found the generalized Chaplygin gas (GCG) which is better fit for latest Supernova data.  After the GCG was introduced, the new model of CG which is called modified CG (MCG) was proposed. An interesting feature of MCG is its ability to explain the evolution of the universe from radiation to $\Lambda$CDM \cite{33,34}. On the  other  hand, it is considered the reconstructing between the scalar field and the  DE models, which is the case, for example,   holographic quintessence \cite{35}, holographic tachyon \cite{36},  interacting new agegraphic tachyon, K-essence and dilaton \cite{37,38}. Meanwhile the simplest explanation of the phantom DE is provided by a scalar field with a negative kinetic energy \cite{39}. Such a field may be motivated from S-brane constructions in string theory \cite{40}. The constraint on parameters   in GCG model,  which is discussed briefly by using the observational data.  Specifically, by inflicting that the energy density of the scalar field must match to the HDE  and the NADE Chaplygin gas density, it was demonstrated that the equation of  fields for the  interacting case  reproduces the equation of field  for  HDE and NADE models. Under such circumstances we  use a measurement of the Hubble parameter as a function of redshift to derive constraints on cosmological parameters. It has also been used to constrain parameters of  HDE and NADE Chaplygin gas  models.  \\
All of these considerations are  mainly investigated in a spatially flat homogeneous and isotropic universe which described by Friedmann-Robertson-Walker (FRW) universe. The theoretical studies and experimental data, which support the existence of an anisotropic phase, lead to consideration the models of universe with anisotropic back ground. Since, the universe is almost isotropic at a large scale, the study of the possible effects of an anisotropic universe in the early time makes the Bianchi  type I (BI) model as a prime alternative for study. Jaffe  \textit{et al}. \cite{41} investigated that removing a Bianchi component from the WMAP data can account for
several large-angle anomalies leaving the universe to be isotropic. Thus the universe may have achieved a slight anisotropic geometry in cosmological
models regardless of the inflation. Further, these models can be classified according to whether anisotropy occurs at an early stage or at later times of
the universe. The models for the early stage can be modified in a way to end inflation with a slight anisotropic geometry \cite{42}.  Very recently, Hossienkhani \cite{43} investigated the interacting ghost DE model with  the quintessence,   tachyon and  K-essence  scalar field in an anisotropic universe. In
\cite{433} by introducing an interacting between DE and DM it was found that the equation of state 
parameter of the interacting DE can cross the phantom line. However, the problem was restricted to the cases that the equation of motion parameter of the universe and anisotropy parameter, are a constant, and the role of time dependence of them  was neglected.\\
Hence, our purpose in this work is to establish a correspondence between the HDE, NADE  and the GCG model. We consider the universe which
has an anisotropic characteristic and we study the effect of time dependence of anisotropy parameter of the BI universe  and reconstruct the potential and the dynamics of the scalar field which describe the Chaplygin cosmology. The paper is organized as following. In section 2 we introduce the general formulation of the field equations in a BI metric. Then we describe the evolution of background cosmology with  generalized Chaplygin gas DE.
 In section 3   we establish the  correspondence between the interacting HDE and the GCG model in BI universe. We reconstruct the potential and the dynamics for the scalar field of the GCG  model, which describe accelerated expansion. In  section 4,  this investigation was extended to the  interacting new agegraphic  GCG DE model.
 In sections 5, 6 we  discuss the $H(z)$ data and the linear evolution of perturbations in HDE and
NADE generalized Chaplygin gas models in BI and compare with the  $\Lambda$CDM  model.
 Eventually we conclude and summarise our results in section 7.

\section{ Reconstruction generalized Chaplygin gas model in anisotropic universe}
To evaluate the influence of both the global expansion and the line of sight conditions on light propagation we
examine an anisotropic accurate solution of the Einstein field equations. The BI cosmology has different expansion rates along the three orthogonal spatial directions, given by the metric
\begin{equation}\label{1}
ds^2=dt^{2}-A^{2}(t)dx^{2}-B^{2}(t)dy^{2}-C^{2}(t)dz^{2},
\end{equation}
where $A(t)$, $B(t)$ and $C(t)$ are the scale factors which describe the anisotropy of the model.  When
$A =B=C$ , the BI model reduces to the flat FRW model. So  BI is the   generalization of the flat FRW model. The non-trivial Christoffel symbols
corresponding to BI universe are
\begin{eqnarray}\label{1a}
&&\Gamma^1_{10}=\frac{\dot{A}}{A},\quad \Gamma^2_{20}=\frac{\dot{B}}{B},\quad\Gamma^3_{30}=\frac{\dot{C}}{C},\cr
&&\Gamma^0_{11}= A\dot{A},\quad\Gamma^0_{22}= B\dot{B} ,\quad\Gamma^0_{33}= C\dot{C},
\end{eqnarray}
where, the aloft dot on the scale factors denote differentiation with respect to time $t$. The energy-momentum tensor is defined as
\begin{eqnarray}\label{2}
T^{\mu}_{\nu}=diag[\rho,-\omega\rho,-\omega\rho,-\omega\rho],
\end{eqnarray}
where $\rho$ and $\omega$ represent the energy density and EoS parameter respectively.  Einstein's field equations  for the  BI metric is given in  (\ref{1}) which lead to the following system of equations \cite{43}
\begin{eqnarray}
&&3H^{2}-\sigma^{2}=\kappa^2(\rho_{m}+\rho_{\Lambda}),   \label{3} \\
&&3H^2+2\dot{H}+\sigma^{2}=-\kappa^2\left(p_{m}+p_{\Lambda}\right), \label{4}\\
&&R=- 6\big( \dot{H} + 2 H^2\big) -2 \sigma^2. \label{5}
\end{eqnarray}
We have taken $\kappa^2=1$,   $\rho_{\Lambda}$ and $p_{\Lambda}$ are  the energy density and pressure of DE, respectively.  Here, we assume that the case
that the shear is dominated comparing with the other
matter fields; $\sigma^{2}\gg 8\pi G\rho_{tot}$. On the other hand, we know
that the shear evolves as $\sigma\propto a^{-3}$. Therefore, from the BI equation, the universe is expanded as $a\propto t^{1/3}$
in the shear dominated epoch. Now we present some important definitions of physical   parameters. The average scale factor $a$,  volume scale factor $V$ and the generalized mean Hubble parameter $H$  are defined as
\begin{eqnarray}\label{7}
a=\sqrt[3]{ABC}, \quad V=ABC, \quad H=\frac{1}{3}(H_1+H_2+H_3),
\end{eqnarray}
where $H_1=\dot{A}/A$, $H_2=\dot{B}/B$ and $H_3=\dot{C}/C$ are defined as the directional Hubble
parameters in the directions of $x$, $y$ and $z$ axis respectively. The expansion scalar $\theta$ and shear scalar $\sigma^2$
are defined as follows
\begin{eqnarray}
\theta&=&3H=u^{u}_{;u}=\frac{\dot{A}}{A}+\frac{\dot{B}}{B}+\frac{\dot{C}}{C},   \label{8} \\
2\sigma^{2}&=&\sigma_{\mu\nu}\sigma^{\mu\nu}=\left(\frac{\dot{A}}{A}\right)^{2}+
\left(\frac{\dot{B}}{B}\right)^{2}+\left(\frac{\dot{C}}{C}\right)^{2} -3H^{2}, \label{9}
\end{eqnarray}
and
\begin{eqnarray}\label{10}
\sigma_{\mu\nu}=\frac{1}{2}\bigg(u_{\mu;\alpha}h_\nu^\alpha+u_{\nu;\alpha}h_\mu^\alpha\bigg)-\frac{1}{3}\theta h_{\mu\nu},
\end{eqnarray}
where $h_{\mu\nu}=g_{\mu\nu}-u_{\mu}u_{\nu}$ defined as the projection tensor. Note that the model  considers pressureless DM $(p_m=0)$.
  The dimensionless density parameters in an anisotropic universe are defined as usual
\begin{eqnarray}\label{11}
\Omega_m =\frac{\rho_m}{\rho_{cr}}, \quad  \Omega_{\Lambda} =\frac{\rho_{\Lambda}}{\rho_{cr}},\quad
\Omega_{\sigma}=\frac{\sigma^2}{3H^2},
\end{eqnarray}
where the critical energy density is $\rho_{cr}=3H^2$.  Equivalently, Eq.  (\ref{3}) can be expressed as
\begin{equation}\label{11a}
H=H_0(\frac{\Omega_{m0}a^{-3}+\Omega_{\sigma 0}a^{-6}}{1-\Omega_\Lambda})^{\frac{1}{2}},
\end{equation}
where $H_0$, $\Omega_{m0}$ and $\Omega_{\sigma 0}$ are the current values for  $H$, $\Omega_{m}$ and $\Omega_{\sigma }$. In the $\Lambda$CDM model Hubble's parameter is $H=H_0(  \Omega_{m0}a^{-3}+\Omega_{\sigma 0}a^{-6} + \Omega_\Lambda)^{\frac{1}{2}}$ and the EoS of  DE is fixed to be $\omega_\Lambda=-1$. Also for model such as $w$CDM (with the constant EoS $w$), it is $H=H_0(  \Omega_{m0}a^{-3}+\Omega_{\sigma 0}a^{-6} + (1- \Omega_{m0}-\Omega_{\sigma 0})a^{-3(1+w)} )^{\frac{1}{2}}$. The currently preferred values of $w$ is given by: $w = -1.01 \pm 0.15$ \cite{431},  $w = -0.98 \pm 0.12$ \cite{551} and $w=-1.13^{+0.24}_{-0.25}$ from  the CMB and baryon acoustic oscillation (BAO) \cite{432}.
 Measuring the effects of DE model in a series of redshift\footnote{Redshift $z=a^{-1}-1$, where high redshift corresponds to early   times.} bins is so necessary to distinguish among the many possibilities. Then,  by using Eq. (\ref{11}), we can rewrite (\ref{3}) in the form of fractional energy densities as
\begin{equation}\label{12}
\Omega_m+\Omega_{\Lambda}=1-\Omega_{\sigma}.
\end{equation}
In the following, we can determine the deceleration parameter $(q)$  as $q=-1-\frac{\dot{H}}{H^2}$.  Comparing Eqs. (\ref{3}) and (\ref{4}), the deceleration parameter is given:
\begin{equation}\label{13}
q=\frac{1}{2}+\frac{3}{2}\frac{p_\Lambda+\Omega_{\sigma}}{\rho+\Omega_{\sigma}},
\end{equation}
where $\rho=\rho_m+\rho_\Lambda$ satisfies the conservation equation,  the DE and DM components do not obey
the energy conservation separately as their interaction. Thus we assume that they respectively satisfy
the following equations of motion,
\begin{eqnarray}\label{14}
\dot{\rho}_{\Lambda}&+&3H\rho_{\Lambda}(1+\omega_{\Lambda})=-Q, \\
\dot{\rho}_m&+&3H\rho_m=Q,\label{15}
\end{eqnarray}
where $Q$ represents the interaction term and we take it as \cite{44}
\begin{eqnarray}\label{16}
Q=3H b^2\rho_\Lambda(1+r),
\end{eqnarray}
where $b^2$ is the coupling constant and  $r=\rho_m/\rho_\Lambda$ is the energy density ratio.\\
Now Let us  consider the case  where the DE is represented by a  generalized  Chaplygin gas (GCG). We have already mentioned that the   GCG was suggested as an alternative model of DE with an exotic EoS, namely \cite{31,45}
\begin{equation}\label{17}
p_{\Lambda}=-\frac{K}{\rho_\Lambda^{\alpha}},
\end{equation}
where $K$ and $0\leq\alpha\leq1$ are the constant (the SCG corresponds to the case $\alpha$= 1).  Eq. (\ref{17}) leads to a density evolution as
\begin{equation}\label{18}
\rho_{\Lambda}=\left(K+D a^{-3\beta}\right)^{\frac{1}{\beta}},
\end{equation}
 where $\beta=1+\alpha$ and  $D=(\frac{c}{a_0^3})^{\beta}$ is a positive integration constant.  In Ref. \cite{46} the energy density of GCG can be derived as $\rho_{GCG}=\rho_0\left(A_s+(1-A_s) a^{-3\beta}\right)^{\frac{1}{\beta}}$ where $A_s=K/\rho_0^{1+\alpha}$. So, the value of $D$ is given by $D=1-K/\rho_0^{1+\alpha}$. This type of matter at the beginning of the cosmological evolution behaves
like dust and at the end of the evolution like a cosmological constant. From Eq. (\ref{18}) it is seen that at the earlier time $\rho_{\Lambda}$ tends to infinite and $\rho_\Lambda=K^{1/\beta}$ at $a\rightarrow\infty$.  In the case of $Da^{-\beta}=-K$, we have $|p|\rightarrow\infty$  in initial time.  At late times, becomes $p_\Lambda=-K^{1/\beta}$ which show that an acceleration universe. Taking derivatives in both sides of  Eq. (\ref{18}) with respect to cosmic time, we obtain
 \begin{figure}
\centerline{ \includegraphics[width=.33\textwidth]{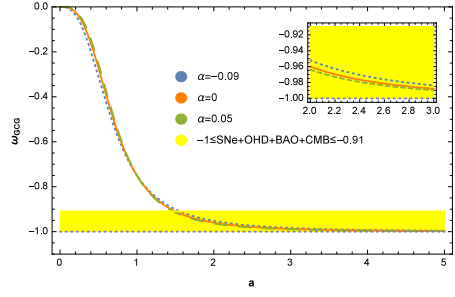}}
\caption {Equation of state parameter of GCG DE versus scale factor $a$, considering    $\Omega_{GCG}=0.956$ and  $\alpha=-0.09$ \cite{46}, $\alpha=0$ and $\alpha=0.05$ \cite{461}.}
 \label{fig:1}
 \end{figure}
\begin{equation}\label{19}
\dot{\rho}_{\Lambda}=-\frac{3DH}{a^{3\beta}}\left(K+D a^{-3\beta}\right)^{-\frac{\alpha}{\beta}}.
\end{equation}
Using Eqs.  (\ref{17})  and (\ref{18}),  the EoS parameter of the GCG model of DE is obtained as
\begin{equation}\label{20}
\omega_{\Lambda}=-1+\frac{Da^{-3\beta}}{K+Da^{-3\beta}}.
\end{equation}
$\omega_{\Lambda 0}=-K/(K+D)$ is the present value of the EoS parameter.
In the following, we consider the cosmology model with  values of parameters:   $\Omega_{GCG}=0.956$ and $\alpha=-0.09$  \cite{46}, $\alpha=0$,  which can be reduced to the standard  DE plus  DM models and $\alpha=0.05$ \cite{461}.  We have plotted the evolution of the EoS parameter of GCG DE  with respect to the scale factor  $a$ in Fig. (1). We see that the EoS parameter translates the universe from matter region towards  vacuum DE region. The   curves representing the GCG are very similar, only the initial slope changes with the change of the $\alpha$ parameter.  Ref. \cite{46}   found that the best fit evolution of $\omega_{GCG}$ is $-1\leq\omega_{GCG}\leq -0.91$  and this result  is consistent with \cite{47}. Now introduce the squared speed of  GCG fluid as
\begin{equation}\label{20a}
v_s^2=\frac{p'_\Lambda}{\rho'_\Lambda}=\omega'_\Lambda\frac{\rho_\Lambda}{\rho'_\Lambda}+\omega_\Lambda,
\end{equation}
which now becomes
\begin{equation}\label{20b}
v_s^2=\frac{K \alpha a^{-3(1+\alpha)}}{D+K \alpha a^{-3(1+\alpha)} }.
\end{equation}
It is found that the model admits a positive squared speed for  $\alpha>0$.  Thus for a stable model we require  $\alpha$ positive.\\
 In the following, we regard the scalar field model as an effective description of an underlying theory of DE with energy density and pressure
\begin{eqnarray}\label{21}
\rho_{\phi}&=&1/2\dot{\phi^2}+V(\phi)=\left(K+D a^{-3\beta}\right)^{\frac{1}{\beta}},\\
p_{\phi}&=&1/2\dot{\phi^2}-V(\phi)=-K\left(K+D a^{-3\beta}\right)^{-\frac{\alpha}{\beta}},\label{22}
\end{eqnarray}
where $\dot{\phi}^2$ and $V(\phi)$ are termed as kinetic energy and scalar potential, respectively.
Now by using Eqs. (\ref{21})  and (\ref{22}) we can easily obtain the   potential and the kinetic energy terms as
\begin{eqnarray}
\dot{\phi^2}&=&\frac{Da^{-3\beta}}{\left(K+D a^{-3\beta}\right)^{\frac{\alpha}{\beta}}}, \label{23}\\
V(\phi)&=&\frac{K+\frac{D}{2}a^{-3\beta}}{\left(K+D a^{-3\beta}\right)^{\frac{\alpha}{\beta}}}.\label{24}
\end{eqnarray}
 The  above equation shows that $\dot{\phi^2}<0$ (giving negative kinetic energy) for  $D a^{-3\beta}<0$. Therefor one can concludes that the scalar field $\phi$ is a phantom field. Efstathiou \textit{et al}. \cite{48} provided the simplest  quintessence models and  obtained   the range of value  $\omega_q$  is $\omega_{q}  \leq -0.6$.  To keep thing  the simple model, then, we shall use a potential  $V\propto q^{-2}$.  As a reference, it is relevant  to mention that long back, Hoyle and Narlikar used C-field (a scalar called creation) with negative kinetic energy for steady state theory of the universe \cite{49}. In the next sections we consider the above equations to determine the potential in the two
cases (i) holographic DE (ii) new agegraphic DE.
 \section{ Correspondence between the interacting holographic   DE  and  generalized Chaplygin gas model  in anisotropy universe}
In this section we consider a  non-isotropic universe. Here  our choice for holographic DE  density is \cite{17}
\begin{equation}\label{25}
\rho_{\Lambda}=\frac{3c^2}{R_h^2},
\end{equation}
$R_h$ is the future event horizon. Suggestive as they are, these
ideas provide no indication about how to pick out the IR cutoff in a
cosmological context.  We are interested in the one proposed in  \cite{18}:
\begin{equation}\label{26}
R_h=a\int^{\infty}_{t}\frac{dt}{a}=a\int^{\infty}_{x}\frac{dx}{aH},
\end{equation}
 where $x = \ln a$.   Note that the  presence of a vacuum energy component makes the above integration confined.  In the case of non-interacting  fluid the conservation equation  for DM can be written as
\begin{equation}\label{27}
\rho_m=\rho_{m0}a^{-3(1+\omega_m)}=\rho_{m0}(1+z)^{3(1+\omega_m)}.
\end{equation}
 We recalls that the reconstruction method is limited to pressureless fluids, so Eq. (\ref{27}) reduces to $\rho_m\propto (1+z)^3$ when dust matter $\omega_m=0$ is assumed.   In \cite{18}, a convenient method to solve  equations is carried out by taking $\Omega_{\Lambda}=\rho_{\Lambda}/\rho_{cr}=c^2/R_h^2H^2$  as the unknown function. The time derivative of the future horizon is given by:
\begin{equation}\label{28}
\dot{R}_h=R_hH-1=\frac{c}{\sqrt{\Omega_{\Lambda}}}-1.
\end{equation}
Taking the derivative with respect to the cosmic time of  (\ref{25}) and using (\ref{28}) we get
\begin{equation}\label{29}
\dot{\rho}_\Lambda=2H\left(\frac{\sqrt{\Omega_{\Lambda}}}{c}-1\right)\rho_{\Lambda}.
\end{equation}
\begin{figure}
\centerline{ \includegraphics[width=.33\textwidth]{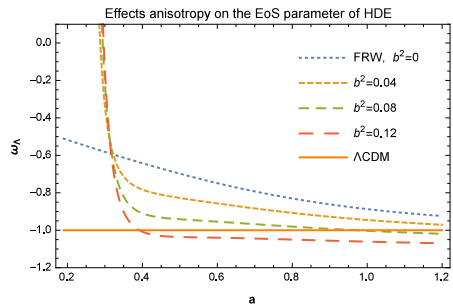}}
\caption {Evolutions of $\omega_\Lambda$ with and without interaction.  The rest of parameter are $c=1$  and  $\Omega_{\sigma0}=0.001 $.}
 \label{fig:2}
 \end{figure}
   \begin{figure}[h]
  \includegraphics[width=.31\textwidth]{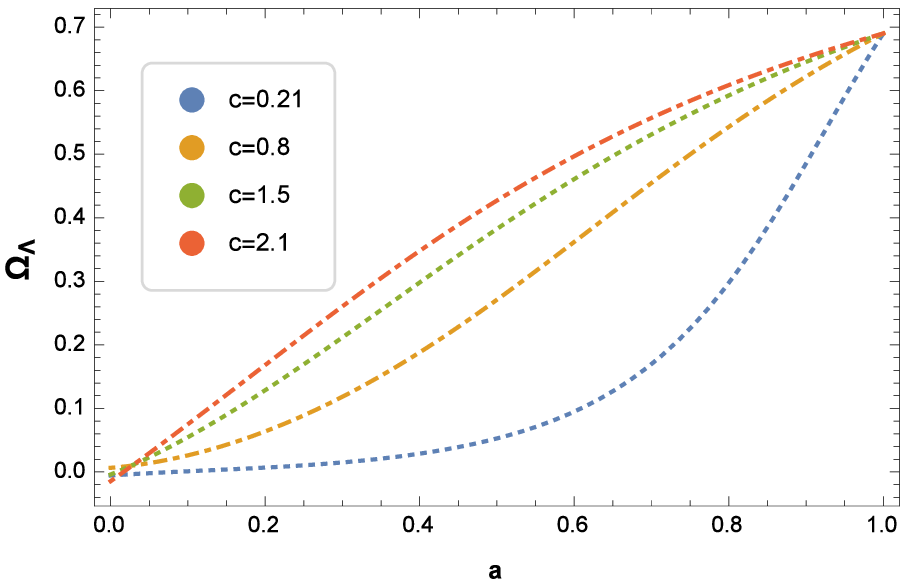}\hspace{1.5cm}
\includegraphics[width=.27\textwidth]{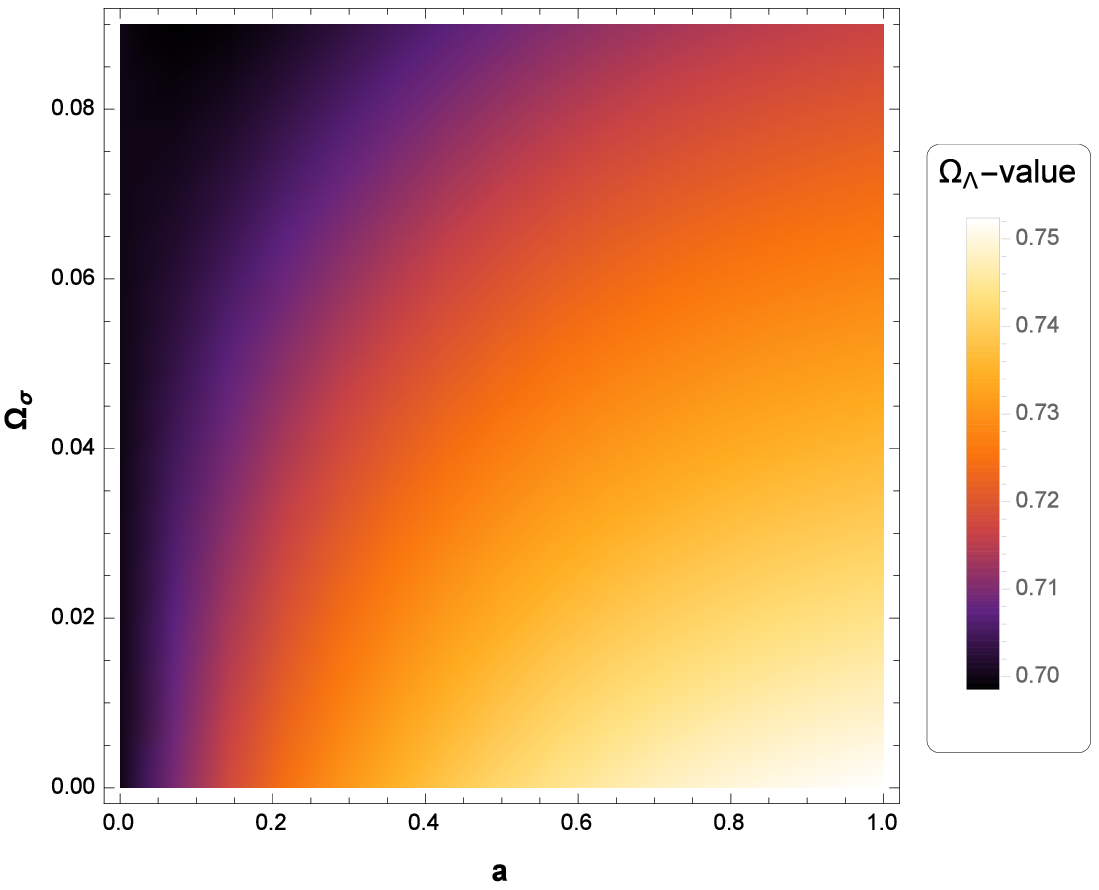}
\caption{ The first  figure representations of   $\Omega_\Lambda$     for   different $c$   and   $\Omega_{\sigma0}=0.001$    while second figure representations of   $\Omega_\Lambda$  versus $\Omega_{\sigma}$ and $a$ for $c=1$. We take for both  $b^2=0.02$.}
 \label{fig:3}
 \end{figure}
 Substituting Eqs. (\ref{16}), (\ref{25}) and (\ref{29}) into (\ref{14})  and using definition $r=(1-\Omega_\sigma-\Omega_\Lambda)/\Omega_\Lambda$, gives the EoS parameter of the interacting HDE model as
\begin{equation}\label{30}
\omega_{\Lambda}=-\frac{1}{3}-\frac{2\sqrt{\Omega_{\Lambda}}}{3c}-\frac{b^2}{\Omega_{\Lambda}}(1-\Omega_{\sigma}).
\end{equation}
 In the far future $(a\rightarrow \infty$, $\Omega_{\sigma}\rightarrow0$ and $\Omega_{\Lambda}\rightarrow1)$, one has $\omega_{\Lambda}=-1/3-2/3c$, so the HDE model does not involve the $\Lambda$CDM model. In the absence of interaction between HDE and CDM, $b^2=0$, using Eq. (\ref{30}), one can see that by considering $c\leq \sqrt{\Omega_{\Lambda}}$. We test this scenario for the interaction between HDE and DM by using some observational results. For the comparison with the phenomenological interacting model, in our scenario the coupling between HDE and DM can be expressed as a counterpart of $b^2$ as in the phenomenological interaction form. In fact, $b^2$ is within the region of the golden supernova data fitting result $b^2=0.00^{+0.11}_{-0.00}$ \cite{50}  and the observed CMB low $l$ data constraint \cite{51}. Now we use 73 SGL data points to estimate $c$ in the model of Markov-Chain Monte Carlo package CosmoMC is $c = 1.9730^{+0.0270}_{-0.8993}$ \cite{511},  the another best-fit   from the strong gravitational lensing (SGL) data is $c = 0.8335^{+0.8031}_{-0.3495}$,  with CBS (CMB+BAO+SN) it is $c =0.6458^{+0.0472}_{-0.0483}$,   the SGL+CBS data is $c = 0.6429^{+0.0515}_{-0.0436}$,  and the $w$CDM model, with the SGL+CBS is $\Omega_{m0} = 0.2891^{+0.0100}_{-0.0092}$ and $ w = -1.0546^{+0.0606}_{-0.0610}$ \cite{512}. \\
  In the numerical calculations, we set $c=1$ and $\Omega_\Lambda^0=0.69$. Figure (2) shows that for $b^2=0$, $\omega_\Lambda$ decreases from $-0.34$ at early times while  for $b^2\neq 0$ it can be observed that the EoS parameter starts from matter dominant  and goes towards lower negative value of   phantom region    for all of the cases of interacting parameter. This behavior arises the  shear scalar evolves as $\sigma^2\propto a^{-6}$.
 Moreover it show that for $b^2> 0.04$, $\omega_\Lambda\sim-1.06$ at present times i.e. $a\rightarrow 1$. This  is  recorder with  the observations \cite{52}. Another best fit  data with the holographic model is $c=0.21$ \cite{521} with SNe Ia, $c=0.7$ \cite{522} with BOOMERANG and WMAP data on the CMB and $c=2.1$ \cite{523} with small $l$ CMB data.  In particular, we have schemed the evolution of $\Omega_\Lambda$ versus scale factor $a$ in an anisotropic universe as shown in figure. (3). In  left panel of  Fig. (3), for  a given $c$,  it is to find that,   $\Omega_\Lambda$ increases  simultaneity when the  $a$ increases; for a given  $a$,  $\Omega_\Lambda$ increases when the  $c$ increases. Finally, figure  (3) (right panel) show the effects of the   anisotropic on the evolutionary behaviour the holographic Chaplygin gas DE model.\\
The main purpose of this work is to investigate correspondence between the Chaplygin gas DE model and the holographic DE model in
the flat anisotropy universe case.  Using Eqs. (\ref{18}), (\ref{20}), (\ref{25}), (\ref{28})  and (\ref{30}), we determine the parameters as
\begin{eqnarray}\label{31}
&&K=(3H^2\Omega_{\Lambda})^{\beta}-Da^{-3\beta},\\
&&D=(3H^2\Omega_{\Lambda}a^3)^{\beta}\left(\frac{2}{3}-\frac{2}{3}\frac{\sqrt{\Omega_{\Lambda}}}{c}-\frac{b^2}{\Omega_\Lambda}(1-\Omega_\sigma)
\right).\label{32}
\end{eqnarray}
Substituting Eq. (\ref{32}) into (\ref{31}) reduces to
\begin{equation}\label{33}
K=(3H^2\Omega_{\Lambda})^{\beta}\left(\frac{1}{3}+\frac{2}{3}\frac{\sqrt{\Omega_{\Lambda}}}{c}+\frac{b^2}{\Omega_\Lambda}(1-\Omega_\sigma)
\right).
\end{equation}
 Now we can rewritten the scalar potential and kinetic energy term as following
\begin{eqnarray}
\dot{\phi^2}&=&2H^2\left(\Omega_{\Lambda}-\frac{\Omega^{\frac{3}{2}}_{\Lambda}}{c}
-\frac{3b^2}{2}(1-\Omega_\sigma)\right), \label{34}\\
V(\phi)&=&H^2\left(2\Omega_{\Lambda}+\frac{\Omega^{\frac{3}{2}}_{\Lambda}}{c}
+\frac{3b^2}{2}(1-\Omega_\sigma)\right).\label{35}
\end{eqnarray}
We now substitute $x=\ln a$, to alter the time derivative into the derivative with logarithm of the scale factor, which is the most useful
function in this case. Consequently from  definition $\dot{\phi}=H \phi'$, one can rewrite  Eq. (\ref{34}) as
\begin{eqnarray}\label{36}
\phi'=\sqrt{2} \left(\Omega_{\Lambda}-\frac{\Omega^{\frac{3}{2}}_{\Lambda}}{c}
-\frac{3b^2}{2}(1-\Omega_\sigma)\right)^{\frac{1}{2}},
\end{eqnarray}
where the prime denotes the differentiation with respect to the  time parameter $x$, Eq.  (\ref{36})  becomes
\begin{eqnarray}\label{37}
\phi(a)-\phi(a_0)=\int _{0}^{ a}\frac{1}{a}\sqrt{2  \left(\Omega_{\Lambda}-\frac{\Omega^{\frac{3}{2}}_{\Lambda}}{c}
-\frac{3b^2}{2}(1-\Omega_\sigma)\right)}da,
\end{eqnarray}
 \begin{figure}[h]
\centerline{ \includegraphics[width=.32\textwidth]{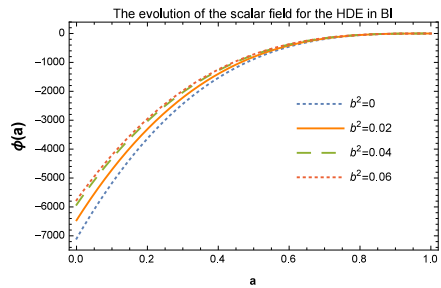}}
\caption{  The evolutionary  scalar field  $\phi$ for the interacting   HDE and GCG  with different  $b^2$.   Auxiliary parameters are  $\Omega^0_{\Lambda}= 0.69$, $\phi(1)=0$, $H_0=72$, $c=1$ and $\Omega_{\sigma0}=0.001$.}
 \label{fig:4}
 \end{figure}
 \begin{figure}[h]
\centerline{ \includegraphics[width=.32\textwidth]{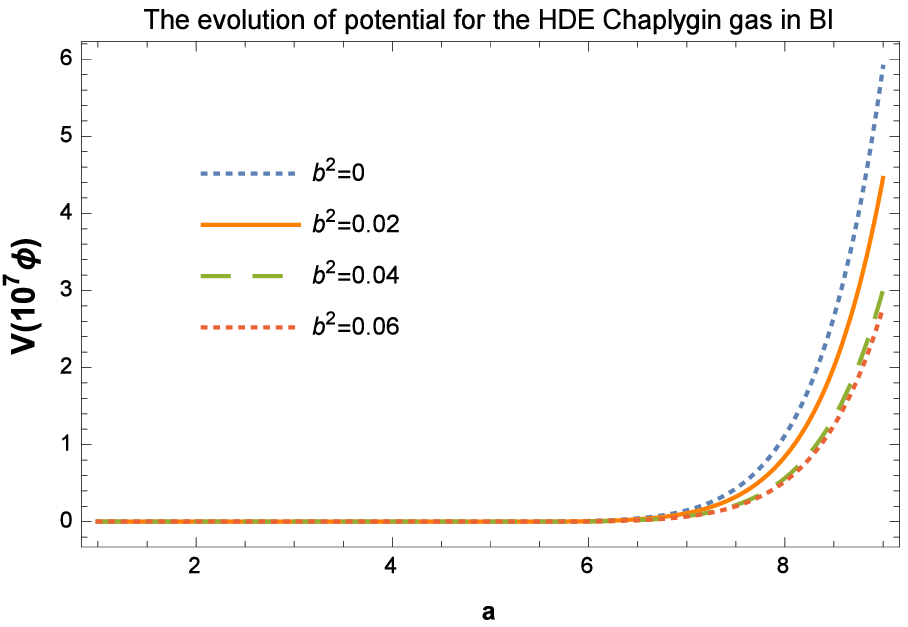}}
\caption{The reconstruction of the potential for  the interacting   HDE and GCG  with different  $b^2$. Auxiliary parameters as in Fig. (4).}
 \label{fig:5}
 \end{figure}
where we take a $a_0 =1$ for the present time, the evolution $\Omega_{\Lambda}$ and $H$ is given by HDE in BI universe \footnote{As one can see in this case the $\Omega_{\Lambda}$ and $H$ can determine  with  the coupling constant $b^2$. In the flat  BI universe case, using   Eqs. (\ref{3}),  (\ref{14}),  (\ref{16}), (\ref{25}),  (\ref{29}) and      (\ref{30}), we can obtain \\
 $\Omega_{\Lambda}'=\Omega_{\Lambda}\bigg(1+3\Omega_\sigma-\Omega_\Lambda+\frac{2}{ c}\sqrt{\Omega_\Lambda}(1-\Omega_\Lambda)-3b^2(1-\Omega_\sigma)\bigg)$  and
 $H'=-\frac{3H}{2}\bigg(1+\Omega_\sigma-\frac{\Omega_\Lambda}{3}-\frac{2}{ 3c}\Omega_\Lambda^{\frac{3}{2}}-b^2(1-\Omega _\sigma)\bigg)$.}.  The evolutionary form of the scaler field and the reconstructed  potential $V(\phi)$ are plotted in Figs. (4) and (5), where again we have taken $\phi(a_0=1)=0$ for the present time. Again, figure (4) shows that  $\phi(a)$ goes up as the scale factor increases here the stronger  interaction is, the slower   $\phi(a)$ which changes as the scale factor increases. Figure (5) illustrate that $V(\phi)$ could increase with the increasing $a$, i.e. the stronger the interaction is, the slower the  $V(\phi)$ varies. Furthermore,
 $V(\phi)$ for the   HDE and GCG without interaction increase faster than that with interaction. To complete, the effective EoS parameter an  anisotropic universe  is obtain as
  \begin{figure}[h]
\includegraphics[width=.3\textwidth]{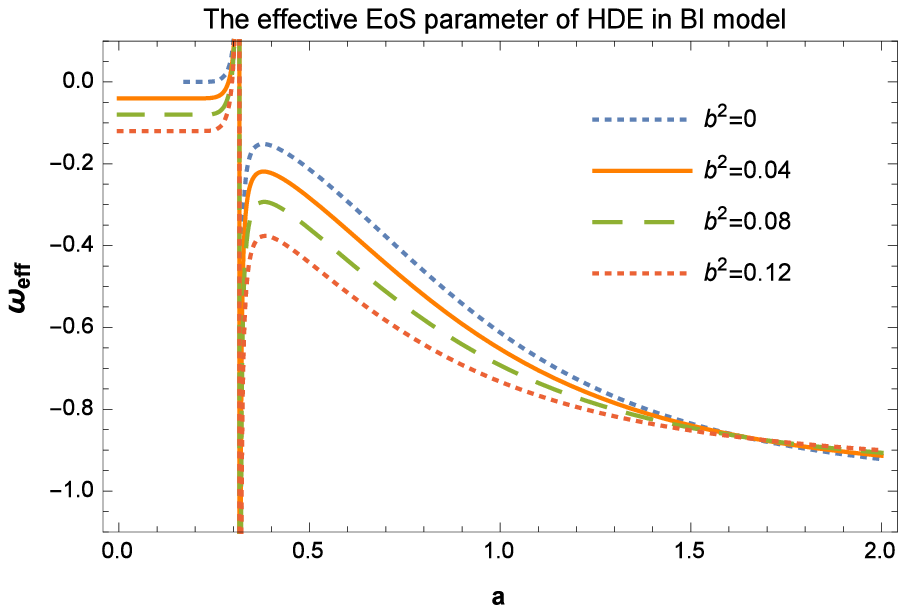}\hspace{1.5cm}
 \includegraphics[width=.3\textwidth]{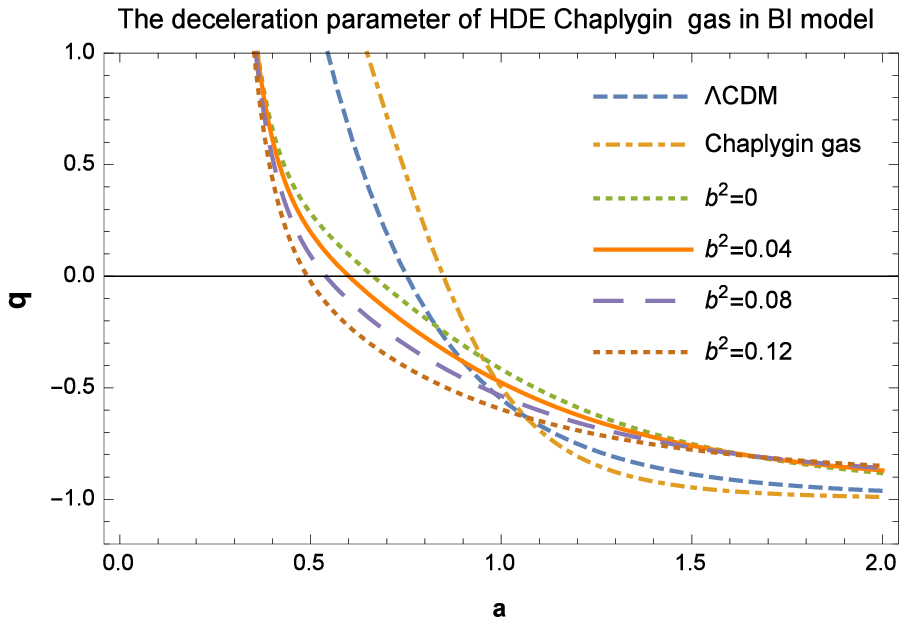}
\caption{The evolutions of $\omega_{eff}$ and $q$  with  scale factor for the interacting HDE with $\Omega^0_{\Lambda}= 0.69$, $c=1$ and $\Omega_{\sigma0}= 0.001$, GCG model  with $A_s=0.7$ and $\alpha=0.02$ and $\Lambda$CDM model with  $\Omega_{m0}=0.3$ and $\Omega^0_{\Lambda}=0.7$.}
 \label{fig:6}
 \end{figure}
\begin{equation}\label{38}
\omega_{eff}=\frac{p_\Lambda}{\rho_m+\rho_\Lambda}=\frac{\Omega_{\Lambda}}{1-\Omega_\sigma}\left(\frac{\frac{1}{2}
\dot{\phi}^{2}-V(\phi)}{\frac{1}{2}\dot{\phi}^{2}+V(\phi)}\right).
\end{equation}
Inserting    Eqs. (\ref{34})  and  (\ref{35}) into (\ref{38}), we obtain
\begin{equation}\label{39}
\omega_{eff}=-\frac{\Omega_\Lambda+\frac{2}{c}\Omega_\Lambda^{\frac{3}{2}}+3b^2(1-\Omega_\sigma)}{3(1-\Omega_\sigma)}.
\end{equation}
With the help   Eqs. (\ref{13}), (\ref{21}), (\ref{22}),  (\ref{32}) and (\ref{33}), we give the deceleration parameter in BI universe
\begin{equation}\label{40}
q=\frac{1}{2}(1-\Omega_\Lambda)+\frac{3}{2}\Omega_\sigma-\frac{\Omega^{\frac{3}{2}}_\Lambda}{c}-\frac{3b^2}{2}(1-\Omega_\sigma).
\end{equation}
 If we take $\Omega_{\Lambda0}=0.69$, $c=1$   and $\Omega_{\sigma0}=0.001$ for now i.e., $a=1$, then Eq. (\ref{39}) gives $\omega_{eff}<-1$  when $b^2>0.08$.
We plot in Fig. (6)  the evolutions of the $\omega_{eff}$ and $q$ of the interacting HDE and GCG with different $b^2$. From left panel of Fig.  6 we see that the  $\omega_{eff}$ of the interacting HDE and GCG cannot  cross the phantom divide at present times. Right panel of Fig. 6   presents that  the universe transitions from a matter dominated epoch at early times to the acceleration phase in the future, as expected. We find that the behaviour of the deceleration parameter for the best-fit universe is quite different from that in the GCG model and  $\Lambda$CDM cosmology.
 In addition, for the case of interacting HDE, we have a cosmic deceleration to acceleration phase at  range of $0.48\leq a\leq0.66$ which is matchable with the observations \cite{53}. For case of $b^2=1.2$, the present value of the best-fit deceleration parameter, $q_0=-0.6$, is significantly smaller than
$q_0=-0.55$ for the $\Lambda$CDM model with $\Omega_{m0}=0.3$ and  also  larger than $q_0=-0.73$ \cite{54}.\\
Now, we  analyze  the model which is using  the    observational  tests:  the differential age of old objects based on the $H(z)$ dependence as well as the data from SGL+CBS and  $w$CDM. The redshift-drift observation, that is called the ``SL test'', is not the only conceptually simple, but also
is a direct probe of cosmic dynamic expansion, although being observationally challenging. In  Ref. \cite{541} introduced the redshift relation  by the  a spectroscopic velocity shift  $\triangle\nu$ as $\triangle\nu\equiv  \triangle z/(1+z)$.  By using the Hubble parameter $H(z)=-\dot{z}/(1+z)$, we obtain \cite{542}
\begin{equation}\label{40a}
\triangle \nu=H_0\triangle t_0\bigg( 1-\frac{\widetilde{H}}{1+z}\bigg),
\end{equation}
where $\widetilde{H}=H/H_0$  and    we have normalized the scale factor to $a(t_0)=1$ and neglected the contribution from relativistic components.
The parameter $\widetilde{H}(z)$ contains all the details of the cosmological model under investigation. It is clear that the function $\widetilde{H}(z)$  is related to the spectroscopic velocity shift via Eq. (\ref{40a}). We will examine the Sandage-Loeb (SL) test, and then examine effects of anisotropy on the HDE and GCG models in   the SL test.
In figure (7) we plot $\triangle \nu$ as function of the source redshift in the flat  BI model case for different values of $\Omega_{\sigma 0}$
assuming a time interval $\triangle t_0=10 $ years for this models. From Fig. (7)  we see that the  interacting holographic and generalized chaplygin gas DE can be distinguished from the SGL+CBS and the $w$CDM models via the
SL test.   In other words, the models shown in Fig. (7) can be easily discriminated using current cosmological tests of the background expansion.
Also we can see that for the case of  $\Omega_{\sigma 0}=0.02$,  $\triangle \nu$   is positive at small redshifts and becomes negative at  $z>0.64$, while for $\Omega_{\sigma 0}\neq0.02$, $\triangle \nu$   is negative in all range of redshift.  Besides,  the amplitude and slope of the signal depend mainly on $\Omega_{\sigma 0}$.
 \begin{figure}[h]
 \includegraphics[width=.3\textwidth]{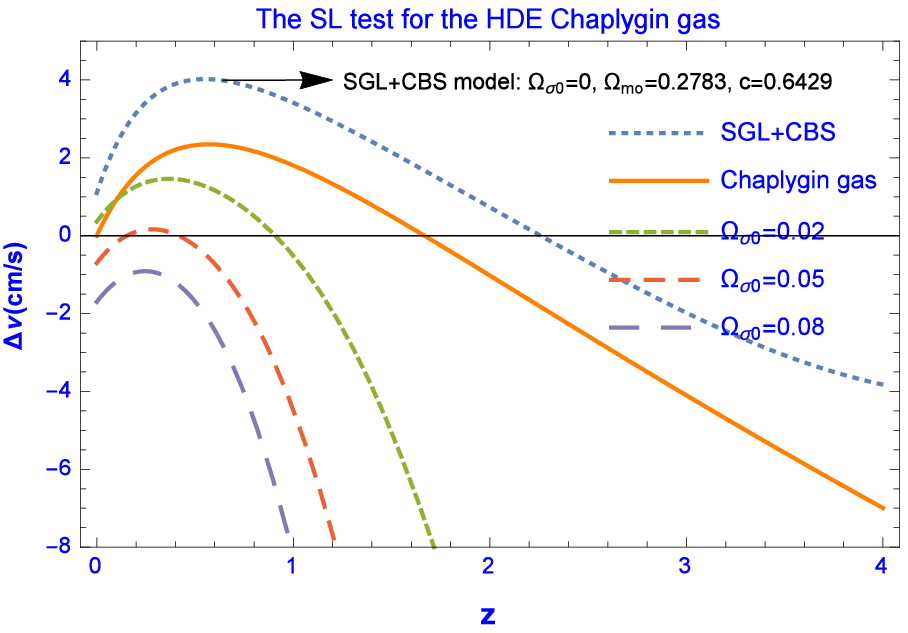}\hspace{1.5cm}
 \includegraphics[width=.3\textwidth]{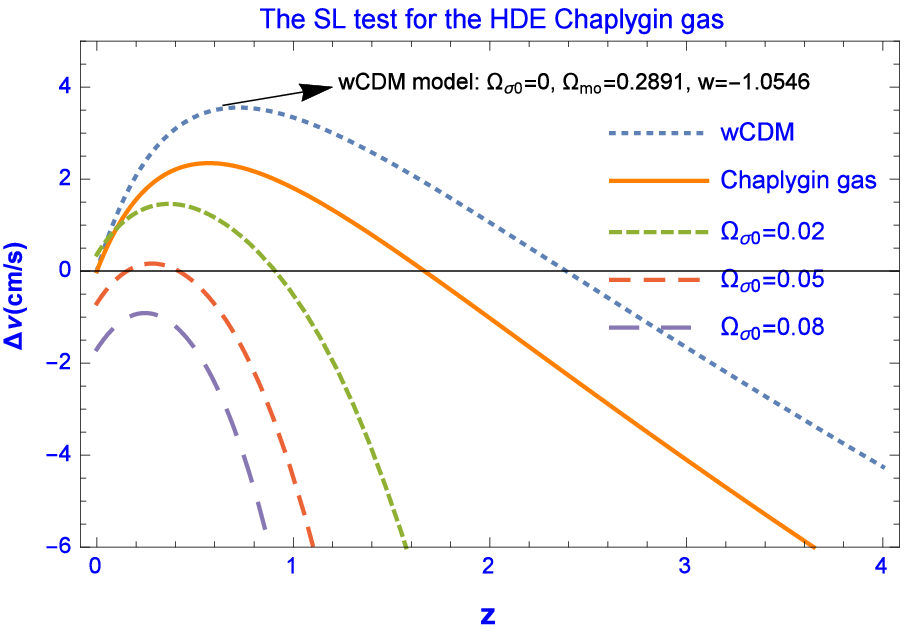}
\caption{The SL test for the HDE  Chaplygin gas model for   different value of     $\Omega_{\sigma 0}$ by comparing  with models as the SGL+CBS model (left panel) and $w$CDM model (right panel). We take for the case of HDE with $\Omega_{m0}=0.28$, $c=1$,  $b^2=0.02$ and $H_0=72~kms^{-1}Mpc^{-1}$  \cite{543} and  GCG model  with $A_s=0.7$ and $\alpha=0.02$.}
 \label{fig:7}
 \end{figure}
\section{ Correspondence between the interacting  new agegraphic DE and Chaplygin gas model of DE in anisotropy universe}
In this section,  we  first review the NADE model. The energy density of the NADE can be written \cite{27}
\begin{equation}\label{41}
 \rho_{\Lambda}=\frac{3n^2}{\eta^2},
\end{equation}
where the conformal time is given by
\begin {equation}\label{42}
 \eta=\int\frac{dt}{a(t)}=\int \frac{da}{Ha^2}.
\end{equation}
If we write $\eta$ to be a definite integral, there will be an integral constant in addition. Thus, we have $\dot{\eta}=1/a$. Now, the fractional energy density of the NADE is given by
\begin{equation}\label{43}
\Omega_{\Lambda}=\frac{n^2}{H^2\eta^2}.
\end{equation}
Taking the derivative of Eq.  (\ref{41}) with respect to the cosmic time and using  (\ref{43}) we  get
\begin{equation}\label{44}
\dot{\rho}_{\Lambda}=-2H \frac{
\sqrt{\Omega_{\Lambda}}}{na}\rho_{\Lambda}.
\end{equation}
Inserting Eq. (\ref{44}) into the continuity equation  (\ref{14}), we obtain the EoS parameter of NADE
\begin{equation}\label{45}
\omega_{\Lambda}=-1+\frac{2\sqrt{\Omega_{\Lambda}}}{3na}-\frac{b^2}{\Omega_\Lambda}(1-\Omega_\sigma).
\end{equation}
 \begin{figure}[h]
\centerline{ \includegraphics[width=.33\textwidth]{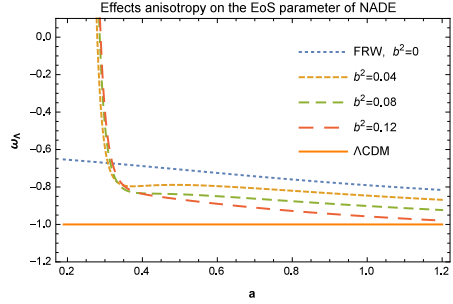}}
\caption { \label{fig:8}Evolutions of  $\omega_\Lambda$  of NADE  with and without interaction. The rest of parameters are $\Omega_{\Lambda}^0=0.69$, $n=2.7$ and $\Omega_{\sigma0}=0.001$.}
 \end{figure}
 \begin{figure}[h]
\includegraphics[width=.31\textwidth]{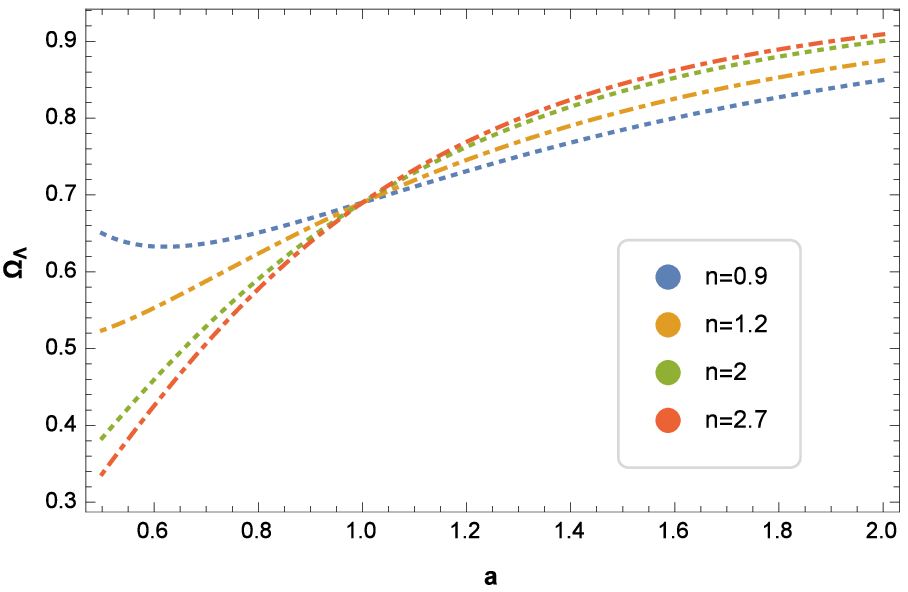}\hspace{1.5cm}
 \includegraphics[width=.27\textwidth]{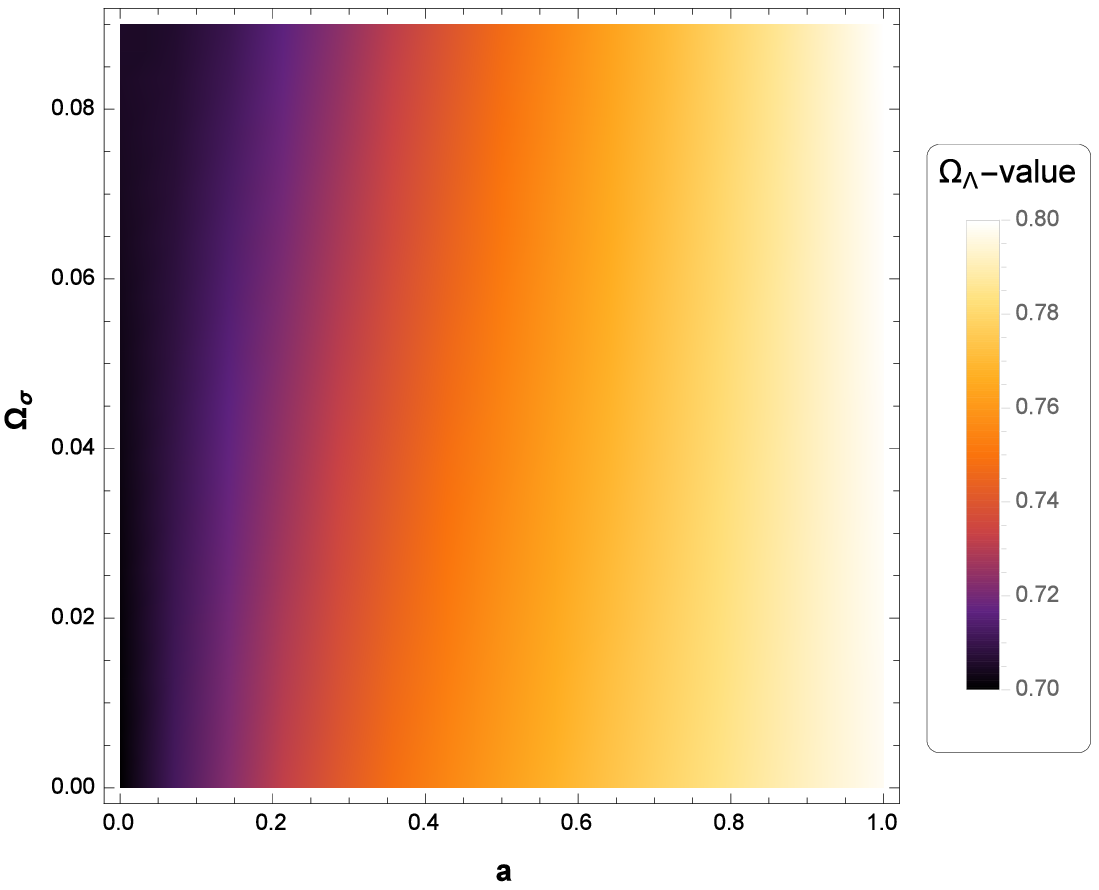}
\caption{ Left panel representations of   $\Omega_\Lambda$   for the NADE with  $a$ for various model parameters $n$   and $\Omega_{\sigma 0}=0.001$  while right panel representations of   $\Omega_\Lambda$ versus  $a$ and  $\Omega_{\sigma }$  for   $n=2.7$. For both  cases, we take  $b^2=0.02$. }
 \label{fig:9}
 \end{figure}
It is important to note that when $b^2=0$, the interacting DE becomes inevitable and Eq. (\ref{45}) reduces to its respective expression in new ADE in general relativity \cite{55}.
In the case  of ($b^2=0$), the present accelerated expansion of our universe can be derived only if
$n>1$ \cite{27}. Note that we take $a=1$ for the present time. In addition, $\omega_\Lambda$ is always larger than
$-1$ and cannot cross the phantom divide $\omega_\Lambda=-1$.   However, in the presence of  the interaction, $b^2\neq 0$,  taking $\Omega_{\Lambda 0}=0.69$, $\Omega_\sigma=0.001$,  $n=2.7$ \cite{28} and  $a=1$ for the present time, Eq. (\ref{45}) gives
\begin{equation}\label{46}
\omega_{\Lambda}=-0.795-1.45b^2.
\end{equation}
It is clear that the phantom EoS $\omega_{\Lambda}<-1$ can be obtained when $ b^2>0.14$  for the coupling between NADE and CDM.
In the late time where $\Omega_\Lambda\rightarrow1$, $\Omega_\sigma\rightarrow0$ and $a\rightarrow \infty$ we have $\omega_\Lambda=-1-b^2$. Thus $\omega_\Lambda <-1$ for $b^2>0$. This implies that in the late time
 $\omega_\Lambda$ necessary crosses the phantom divide in the presence interacting DM and DE.
In the numerical calculations, we set $n=2.7$, $\Omega_\Lambda^0=0.69$ and $\Omega_{\sigma0}=0.001$. From Fig. (8) we   see that  for  $b^2\neq0$, $\omega_\Lambda$ decreases from matter dominant  at early times  while for  $b^2=0$ (FRW), $\omega_\Lambda$ decreases  with  the  $a$ increase  and  its  less steep compared to an interaction term at late times. We see that  for $b^2=1.2$, $\omega_{\Lambda 0}=-0.96$ at present time.  Therefore the EoS parameter is consistent with the WMAP observation \cite{551}.  In figure (9) (left panel), we plot the evolution of the density parameter $\Omega_\Lambda$ for  $b^2=0.02$   as a  function of the $a$ for  different value of   $n$. Moreover, we can see that at the   early time $\Omega_{\Lambda}$ decreases with the increase of $n$, while increases with the increase of $n$ when $a>0$. Also the anisotropy effects are clearly seen in right panel of figure (9).  So   the $\Omega_{\Lambda}$ decreases slowly with increasing of $\Omega_{\sigma}$.  This is consistent with Eq.  (\ref{12}).\\
Next, we suggest a correspondence between the new agegraphic  DE  scenario and the generalized Chaplygin gas DE model.  To do this, comparing Eqs. (\ref{45}), (\ref{20}) and using (\ref{31}), we reach
\begin{equation}\label{47}
K=(3H^2\Omega_{\Lambda})^{\beta}\left(1-\frac{2\sqrt{\Omega_{\Lambda}}}{3na}+\frac{b^2}{\Omega_\Lambda}(1-\Omega_\sigma)\right),
\end{equation}
and
\begin{equation}\label{48}
D=(3H^2\Omega_{\Lambda}a^3)^{\beta}\left(\frac{2\sqrt{\Omega_{\Lambda}}}{3na}-\frac{b^2}{\Omega_\Lambda}(1-\Omega_\sigma)\right).
\end{equation}
We  reconstruct the  kinetic energy  and   scalar potential term as
\begin{eqnarray}
\dot{\phi^2}&=&H^2\left(-3b^2(1-\Omega_\sigma)+\frac{2\Omega^{\frac{3}{2}}_{\Lambda}}{na}\right), \label{49}\\
V(\phi)&=&H^2\left(3\Omega_{\Lambda}+\frac{3b^2}{2}(1-\Omega_\sigma)-\frac{\Omega^{\frac{3}{2}}_{\Lambda}}{na}\right).\label{50}
\end{eqnarray}
Now since  definition $\dot{\phi}=H\phi'$, we get
\begin{eqnarray}\label{51}
\phi(a)-\phi(a_0)=\int _{0}^{ a}\frac{1}{a}\sqrt{-3b^2(1-\Omega_\sigma)+\frac{2\Omega^{\frac{3}{2}}_{\Lambda}}{na}}da,
\end{eqnarray}
\begin{figure}
\centerline{ \includegraphics[width=0.33\textwidth]{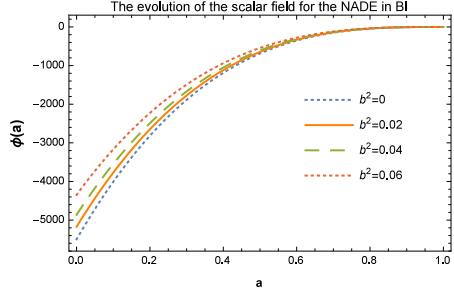}}
\caption{{\small\label{fig:10}The evolutionary  scalar field  $\phi$ for the interacting   NADE and GCG  with different  $b^2$ considering  $\Omega_\Lambda^0=0.69$, $\Omega_{\sigma0}=0.001$,  $n=2.7$ and $\phi(1)=0$.}}
\end{figure}
\begin{figure}
\centerline{ \includegraphics[width=0.33\textwidth]{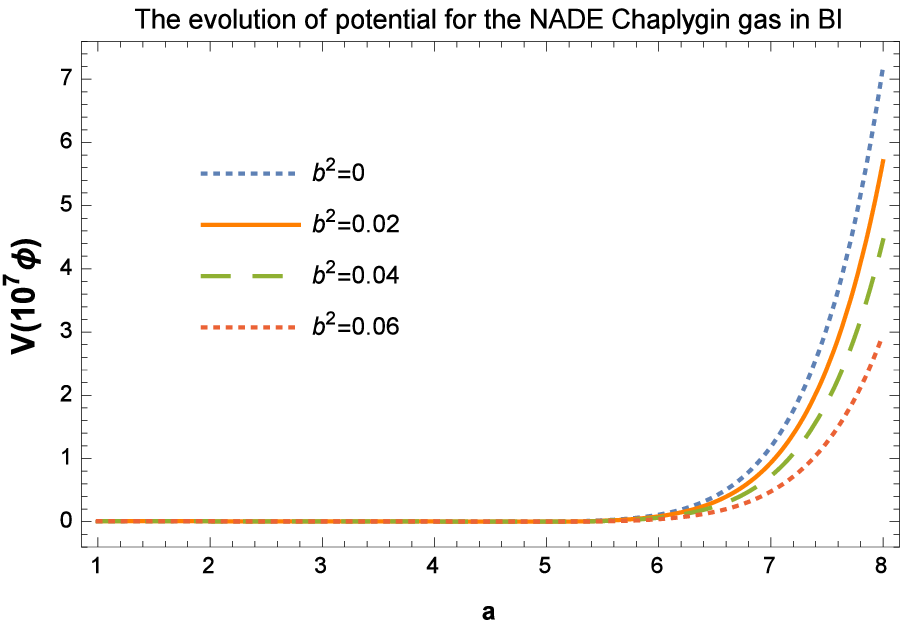}}
\caption{{\small\label{fig:11}The reconstruction of the potential for  the interacting   NADE and GCG  with different  $b^2$.  Auxiliary parameters as in Fig. (10).}}
\end{figure}
where $a_0$ is the present time value of the scale factor,   $\Omega_{\Lambda}$ and $H$ is given by NADE in BI universe \footnote{Taking the derivative of both side of the BI equation (\ref{3})  with respect to the cosmic time, and using    Eqs.   (\ref{12}), (\ref{14}),  (\ref{16}), (\ref{41}),  (\ref{43}) and      (\ref{45}), we can obtain $\Omega_{\Lambda}$ and $H$, respectively,
 $\Omega_{\Lambda}'=3\Omega_{\Lambda}\left(\Omega_\sigma+(1-\Omega_\Lambda)(1-\frac{2}{3na}\sqrt{\Omega_\Lambda})-b^2(1-\Omega_\sigma)\right)$  and \\ $\frac{H'}{H}=-\frac{3}{2}(1-\Omega_\Lambda+\Omega_\sigma)-\frac{\Omega_\Lambda^{\frac{3}{2}}}{na}+\frac{3}{2}b^2(1-\Omega_\sigma)$.}.
Therefore, we have established an interacting new agegraphic and  generalized Chaplygin  gas  DE model and
reconstructed the potential and the dynamics of scalar field in an anisotropic universe.
The evolution of the   scalar filed, Eq. (\ref{51}), for three different values of $b^2$ is plotted in Fig. (10). Figure (10) shows that the scalar field
increases (and hence the kinetic energy $\dot{\phi}^2$ of the potential increases) with the passage of time.
The potential $V(\phi)$ versus $a$  for three different value of the  $b^2$  are shown in figure (11), indicating the increasing behavior.  Also, we can see that at the initial time there is no difference between various values of $b^2$. After that, increasing $b^2$ decreases the value of $V(\phi)$.\\
 Inserting Eqs. (\ref{49}) and (\ref{50}) into (\ref{38}), we obtain  the effective EoS parameter an anisotropic universe
\begin{equation}\label{52}
\omega_{eff}=-\frac{\Omega_\Lambda}{1-\Omega_\sigma}+\frac{2}{3na}(\frac{\Omega^{\frac{3}{2}}_{\Lambda}}{1-\Omega_\sigma})-b^2.
\end{equation}
 Finally, we give the deceleration parameter of interacting NADE and GCG in BI universe
\begin{equation}\label{53}
q=\frac{1}{2}(1-3\Omega_\Lambda)+\frac{3}{2}\Omega_{\sigma}+\frac{\Omega^{\frac{3}{2}}_{\Lambda}}{na}-\frac{3}{2}b^2(1-\Omega_\sigma).
\end{equation}
\begin{figure}
\includegraphics[width=.3\textwidth]{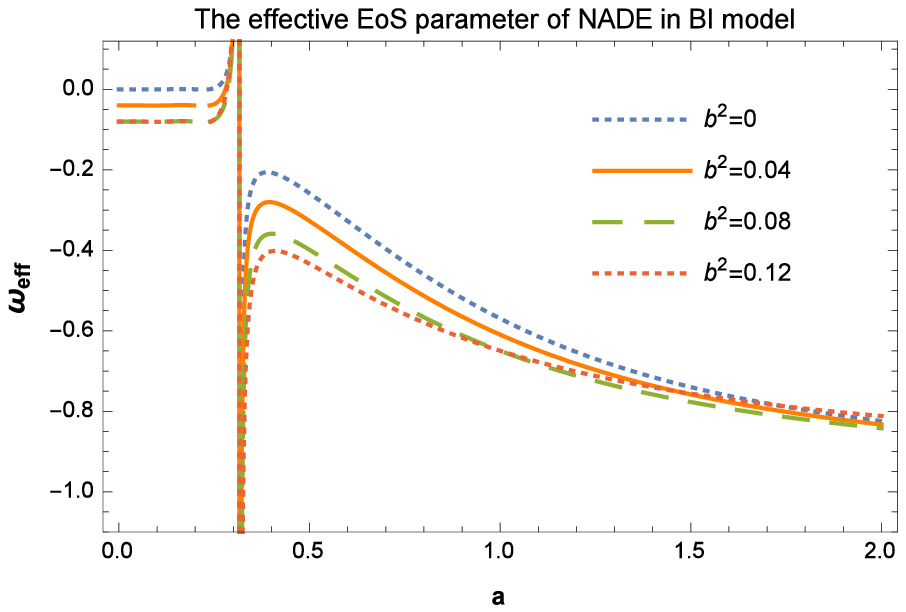}\hspace{1.5cm}
 \includegraphics[width=.3\textwidth]{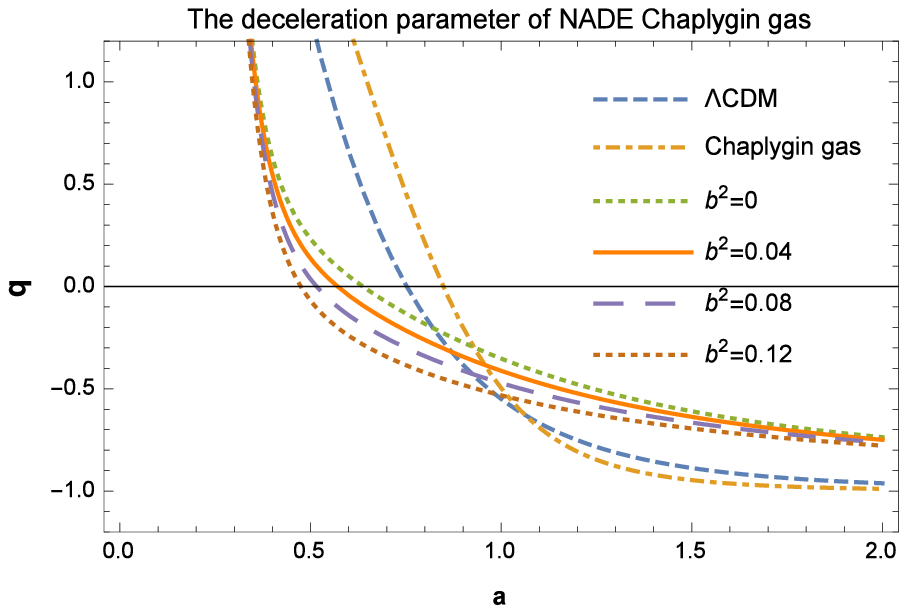}
\caption{The evolutions of $\omega_{eff}$ and $q$  with  scale factor for the interacting NADE with $\Omega^0_{\Lambda}= 0.69$, $n=2.7$ and $\Omega_{\sigma0}= 0.001$, GCG model  with $A_s=0.7$ and $\alpha=0.02$ and $\Lambda$CDM model with  $\Omega_{m0}=0.3$ and $\Omega^0_{\Lambda}=0.7$.}
 \label{fig:12}
 \end{figure}
If we take $\Omega_{\Lambda0}=0.69$,  $\Omega_{\sigma}=0.001$, $n=2.7$ \cite{28} and  $a=1$ for the present time, then Eq. (\ref{52}) will give $\omega_{eff}<-1$ when $b^2>0.45$.  This mentions that the EoS parameter has a phantom behavior.  The evolution of the   effective EoS and deceleration parameter  is plotted in Fig. (12). From left panel of Fig. (12) we see that  for all of the cases of interacting parameter, $\omega_{eff}$ of the NADE cannot have a transition from $\omega_{eff}<-1$. Recent studies have constructed  $q(z)$ takeing into account that the strongest evidence of accelerations happens at redshift of  $z\sim0.2$. In order to do so, the researcher  have  set $q(z)=1/2 (q_1 z+q_2)/(1+z)^2$ to reconstruct it and after that they have obtained  $q(z)\sim-0.31$ by fitting this model  to the  observational data \cite{56,57}.  Also it found that $q<0$ for $0\leqslant z \leqslant 0.2$ within the  $ 3\sigma$ level.
Under such circumstances    and  considering the Eq. (\ref{53}), the present value of the deceleration parameter for the interacting   NADE   in BI models with $b^2=0.12$   is $q_0\sim-0.53$ which is consistent with observations \cite{58}. Moreover,  for the case of interacting NADE, transition from
deceleration to acceleration occurs at rang of  $0.47\leq a\leq0.63$. For the flat $\Lambda$CDM model,   the deceleration parameter $q$ passes the transition point at
$a=0.56$ \cite{581}.   Eventually, the universe will undergo accelerated expansion at the late time forever and cannot come back to decelerated
expansion, as shown in Fig. (12).  These behaviors are similar Refs. \cite{27,28}.  In addition to, the fall of $q$ with scale factor is much steeper in the case GCG and $\Lambda$CDM models  in compare with interacting NADE model. \\
Finally, we examine the Sandage-Loeb (SL) test, and then examine effects of anisotropy on the NADE Chaplygin gas in the SL test. To do this using Eq. (\ref{40a})  we can obtain
\begin{equation}\label{54}
\triangle \nu=H_0\triangle t_0\bigg( 1-(1+z)^{-1}(\frac{\Omega_{m0}(1+z)^{ 3}+\Omega_{\sigma 0}(1+z)^{ 6}}{1-\Omega_\Lambda})^{\frac{1}{2}}\bigg),
\end{equation}
where we set $\triangle t_0=10 $ years and $\Omega_\Lambda$ is given by (\ref{43}).  We reconstruct the velocity shift
behavior in the NADE Chaplygin gas  model respect $z$ for different value of the $\Omega_{\sigma 0}$ in Fig . (13). We have chosen the fractional matter density $\Omega_{m0}=0.274$ from $\Lambda$CDM  \cite{59} and $n = 2.807$ \cite{60}.
From Fig. (13) we see that the new agegraphic and generalized Chaplygin gas  DE  in BI model can be distinguished from the $\Lambda$CDM  model via the SL test. \\
\begin{figure}
\centerline{ \includegraphics[width=0.33\textwidth]{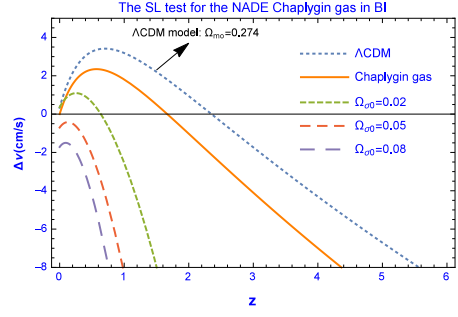}}
\caption{{\small\label{fig:13}The SL test for the NADE Chaplygin gas model for   different value of  the anisotropy energy density parameter   $\Omega_{\sigma0}$ and  the $\Lambda$CDM  model. We take    $\Omega_{m0}=0.28$, $b^2=0.02$ and  $n=2.8$.}}
\end{figure}
\section{Cosmological evolution of the Hubble parameter  of  different dark energy in BI universe and comparison with the $\Lambda$CDM model}
In the section, we   further compare the  expansion rate $H(z)$  with that predicted by  different  models i.e.,  HDE, NADE, GCG and $\Lambda$CDM.  As recently proposed by \cite{61}, these can be used to determine  $H(z)=-\frac{1}{1+z}\frac{dz}{z}$.  Therefore a determination of  $dz/dt$ directly measures  $H(z)$. In \cite{61} it was demonstrated the feasibility of the method by applying it to a $z\sim 0$ sample.  For
the comparison with the phenomenological interacting model, in our scenario the coupling between
HDE, NADE, GCG  and DM can be expressed by $b^2$ parameter as in the phenomenological interaction form. The constraint results from CMB and BAO presented that the mean
values of interaction rate were $b^2= -0.61^{+0.12}_{-0.25}$ from CMB and BAO measurements \cite{62},  $b^2= -0.67^{+0.086}_{-0.17}$ from CMB
and Hubble Space Telescope (HST) tests \cite{63}, and $b^2= 0.00328^{+0.000736+0.00549+0.00816}_{-0.00328-0.00328-0.00328}$ from the redshift space distortion (RSD) date \cite{64}. Fig. (14) shows the comparison of the $H(z)$ estimates with different cosmological models for the cases of $H_0=72~kms^{-1}Mpc^{-1}$, $\Omega_{\sigma0}=0.001$ and $b^2=0.02$ with an anisotropic universe.  The values of $H(z)$   are fully compatible with $\Lambda$CDM, constraining the expansion rate very firmly.  The redshift range $0.5 < z < 2$ is critical to
disentangle many different cosmologies, as can be seen from Fig. (14).  It is shown that in a BI model although HDE model   performs a little poorer than $\Lambda$CDM model, but it performs better than NADE and GCG models. So, among these three   DE models, HDE in BI is more favored by the observational data.  Again we plot the $H(z)$ and effects of anisotropy on    both the HDE and NADE as   shown in figure (15). From Fig. (15), we can clearly see that
for different $\Omega_{\sigma 0}$ parameter value the process of cosmic evolution looks quite similar, i.e., the bigger value the $\Omega_{\sigma 0}$ parameter is
taken, the best value the Hubble expansion rate $H(z)$ is gotten. This implies that the BI model would play a more important role for constraining
the models with more parameters.
\begin{figure}
\centerline{ \includegraphics[width=0.3\textwidth]{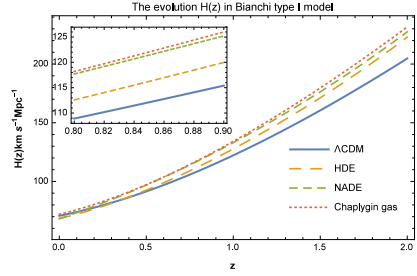}}
\caption{{\small\label{fig:14}The value of the Hubble parameter as a function of redshift as derived from the four models   with $\Omega_{\sigma 0}=0.001$  $c=0.818$,  $n=2.807$,   $A_s=0.7$,  $\alpha=0.02$, $\Omega_{m0}=0.277$ \cite{60} and $H_0=72~kms^{-1}Mpc^{-1}$ \cite{543}. }}
 \end{figure}
\begin{figure}
\includegraphics[width=.3\textwidth]{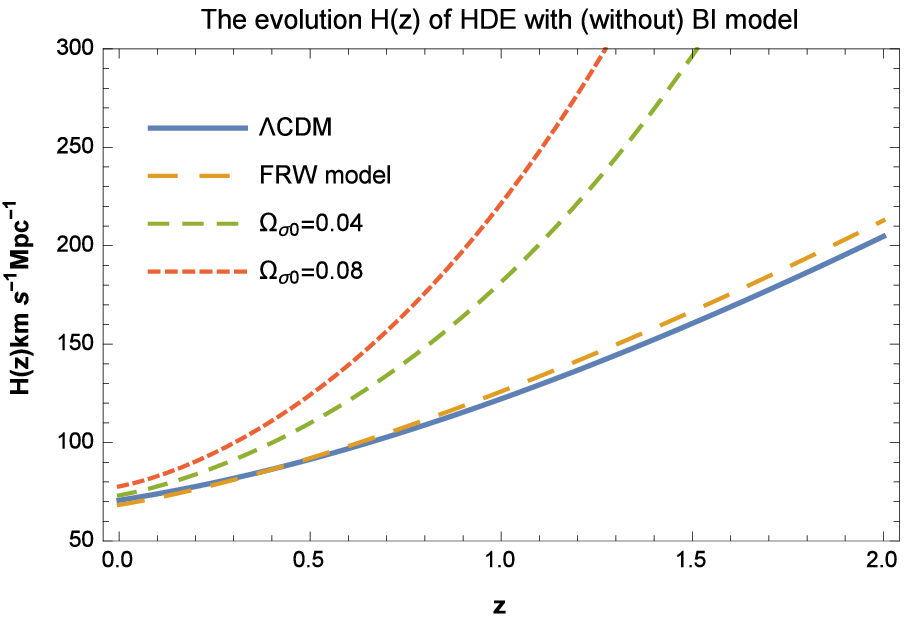}\hspace{1.5cm}
 \includegraphics[width=.3\textwidth]{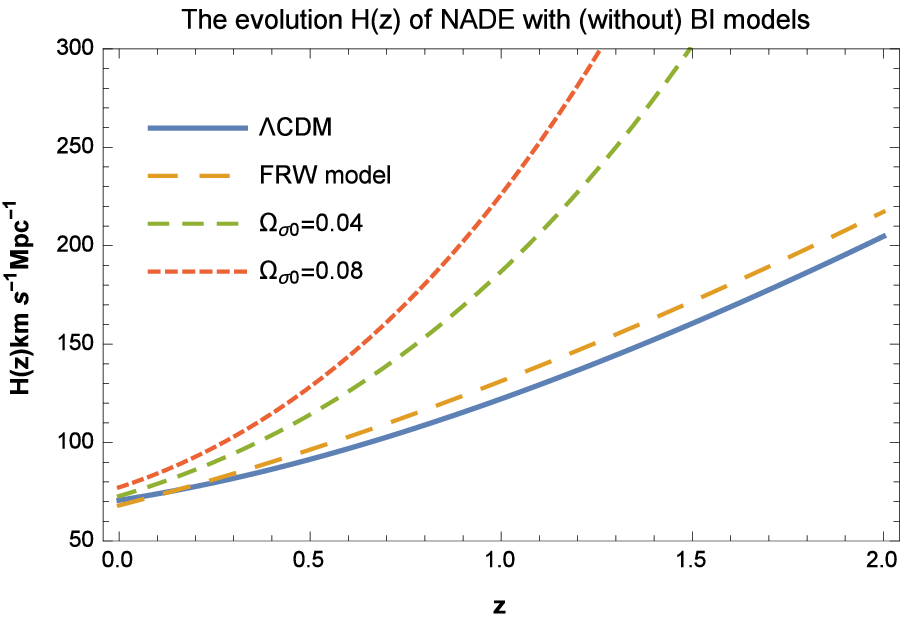}
\caption{The evolution of $H(z)$ versus redshift $z$  of both the HDE and  NADE  for different values of    parameter     $\Omega_{\sigma 0}$.   The rest of parameters are the same as for Fig. (14). }
 \label{fig:15}
 \end{figure}
\section{Linear perturbation theory in anisotropic universe  }
\begin{figure}[h]
 \includegraphics[width=.3\textwidth]{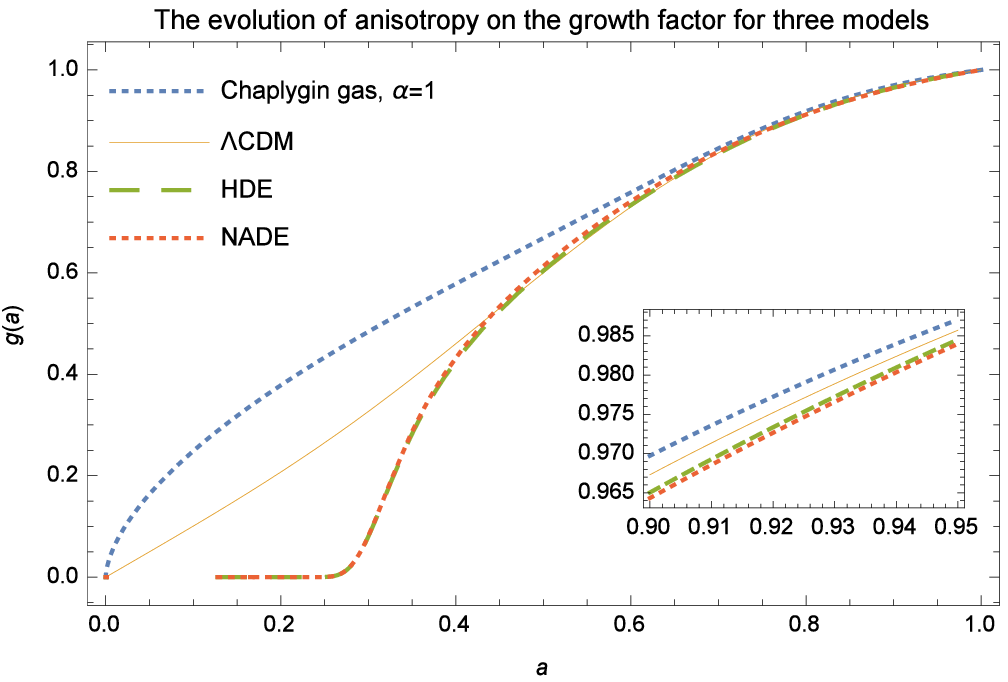}\hspace{.5cm}
\includegraphics[width=.3\textwidth]{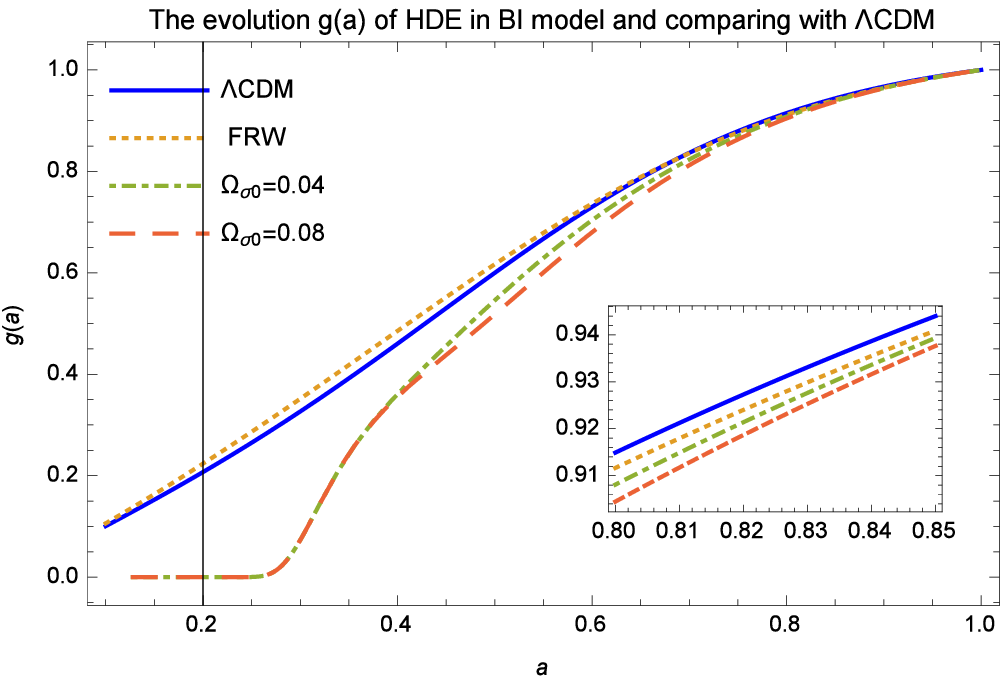}\hspace{.5cm}
\includegraphics[width=.3\textwidth]{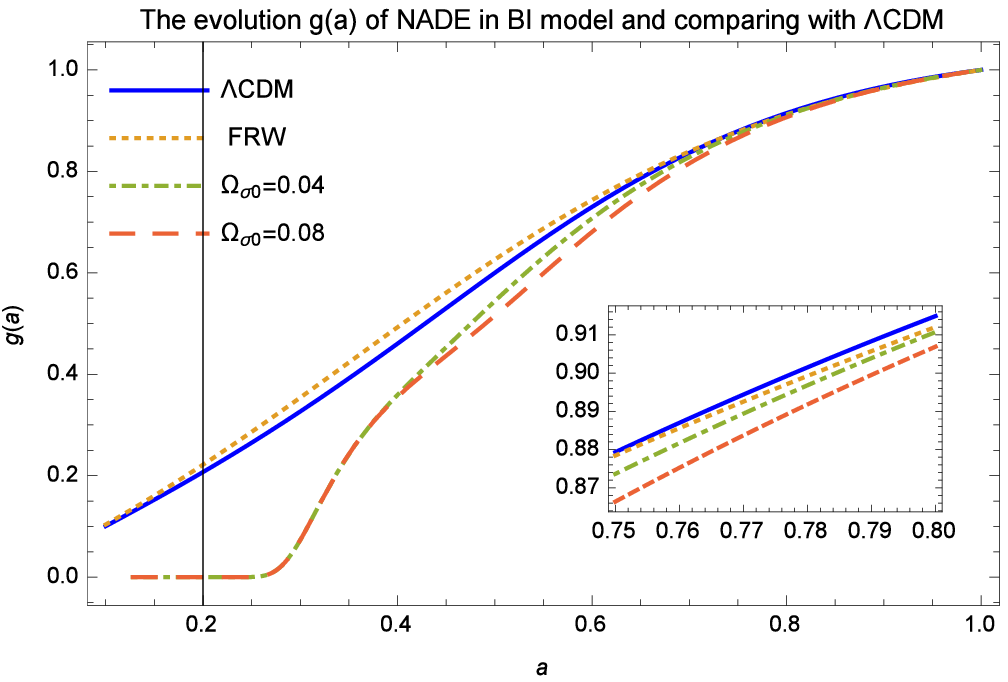}
\caption{Right panel: Time evolution of the growth factor as a function of the scale factor for the three cosmological models in an anisotropic universe. To compare the three models, we thus fix the variance for the HDE and NADE model ($\Omega_{\sigma0}=0.001$). Middle panel: Time evolution of the growth factor for  different value of the anisotropy energy density parameter $\Omega_{\sigma0}$ and comparing to the $\Lambda$CDM and FRW in HDE models with $c=1$.  Left panel: Same as middle panel for NADE with $n=2.7$. }
 \label{fig:16}
 \end{figure}
 Finally, we discuss the linear perturbation theory of non-relativistic dust matter, $g(a)$, for the different DE  models and compare it with the solution
found for the $\Lambda$CDM and FRW models. The differential equation for the evolution of the growth factor
$g(a)$ is given by  \cite{66,67}
 \begin{equation}\label{55}
g''(a)+(\frac{3}{a}+\frac{E'(a)}{E(a)})g'(a)-\frac{3}{2}\frac{\Omega_{m0}}{a^5E^2(a)}g(a)=0,
\end{equation}
where $E=H/H_0$.  For a non interacting  DE model, we solve numerically Eq. (\ref{55}) for studying the linear growth with  four DE models  in BI. Then, we compare the linear growth in   the HDE and NADE  generalized  Chaplygin gas    models with the linear growths in  the $\Lambda$CDM and FRW models.
 To evaluate the initial conditions, since we are in the linear regime, we take that the linear growth factor has a power law solution,
$g(a)\propto  a^n$, with $n>1$, then the  linear growth should grow in time. In Fig. (16) we show the growth factor by the scale factor $a$  for the three DE models considered in this work, as compared to the $\Lambda$CDM and FRW models.  The left panel show that  the growth factor in the GCG  model is larger than the three DE models seen in this work. But  for  small scale factors, the growth factor in the $\Lambda$CDM model is larger than those of HDE and NADE models, while for   the range of $0.49<a<0.73$, it  becomes smaller than those of HDE and NADE models, then it is again greater than the  that of  HDE and NADE models. This means that, at the beginning, the  growth factor in anisotropy for DE models is zero and the $\Lambda$CDM is  more efficient than HDE and NADE models. In both the middle and right figure (16), we see that in the FRW model, the growth
factor evolves proportionally to the scale factor, as expected.  For the $\Lambda$CDM model, we notice
that the evolution of $g(a)$ evolves more slowly than in the FRW case. In the cases of HDE and NADE models with $\Omega_{\sigma0}\neq 0$ (anisotropic universe), $g(a)$    is smaller even when compared to the $\Lambda$CDM model. However, for rather larger scale factors, the growth factor in the  FRW universe becomes smaller than the  $\Lambda$CDM model while it is still larger enough than that of HDE and NADE models in an anisotropic universe. This result is consistent with Ref. \cite{68}.

\section{Conclusion}
We have considered a correspondence between the interaction of HDE and NADE scenarios and the Chaplygin gas model of DE in an anisotropic universe. In particular, we reconstructed the field equations of DE model  in an anisotropic universe. The Chaplygin gas model plays a very crucial role in the EoS fluid description of DE in   cosmology.    The constraints on the GCG model  are given by using   observations of SNe+OHD+BAO+CMB \cite{46,461,47}. The only parameter in this model which needs to be fitted by observational data is the parameter $\alpha=-0.09,0,0.05$.  Furthermore, it is shown that  the GCG model in BI
can drive the universe from a  matter dominated phase to an accelerated expansion phase, behaving like   matter in early times and as  vacuum DE region i.e.,  $\omega_{GCG}\rightarrow-1$ at  late times, which it consistent with the observational data  \cite{46,461,47}. Then we have described this  ``GCG"   as BI universe having a scalar field and found its self-interacting potential.
In what follows,  we have presented the evolution of GCG models for both the HDE and NADE depending on the values of parameters. For the case of HDE dominated universe, i.e., $\Omega_{\Lambda}\sim1$; if we consider $c>1$ and $b^2=0$ (non interaction) then the expansion will in quintessence regime, while for $c<1$, phantom evolution of the universe can be observed. Besides,  it can be observed that for selected value $b^2>0.04$, the EoS parameter   can cross  the phantom region and   $\omega_\Lambda\sim-1.06$ at present times which  the model has agreement with  Ref. \cite{52} (see Fig. (2)). But in case of NADE having $\Omega_{\Lambda}\sim1$, shows that EoS parameter  can be less than $-1$ if  $b^2=0$ and  $n<0$ but observational points of view propose $n=2.76^{+0.11130}_{-0.109}$ \cite{27,28} which permits the phantom era.
We   also reconstructed the dynamics and the potential of the Chaplygin
gas scalar field according the evolution of both the interacting HDE  and NADE models which can describe the
phantomic accelerated expansion of the BI universe. To do that the  holographic and new agegraphic Chaplygin gas scalar field for a given $b^2$ increases
with increasing the scale factor.  Also for a given scale factor, it increases with increasing  $b^2$. The  holographic and new agegraphic Chaplygin  potential $V(\phi)$ for a given $b^2$, increases with increasing the scalar filed. For a given scalar field,
$V(\phi)$ decreases with increasing $b^2$. These results have been shown in figures (4), (5), (10) and (11).
On the basis of the above considerations, it seems reasonable to investigate an  anisotropic universe, in which the present cosmic acceleration
is followed by a decelerated expansion   in  an early matter dominant phase. In other words,    it indicates that the values of transition scale factor and current deceleration parameter are $a\sim 0.84$ and $q_0=-0.49$  for the case of generalized Chaplygin gas,   $0.48\leq a\leq0.66$ and $q_{0}=-0.6$ for the case of holographic   DE with $b^2=1.2$ and    $0.47\leq a\leq0.63$, $q_{0}=-0.53$ for new agegraphic    DE model while for the case of  $\Lambda$CDM model,   the deceleration parameter   passes the transition point at $a=0.56$ \cite{581}. This description is  allows for an unambiguous confrontation with observational data. For this purpose, several studies were performed aiming to constrain the parameter space of the model using observations data. In particular, the   holographic and new agegraphic DE   and GCG models was explored with the SL test in BI model. In order words, the best way to probe models with such  interaction between DM and DE  is to map out cosmic expansion during
the matter dominated phase. The SL tests offers a unique tool to do just that. So, the SL test can be used to distinguish the  HDE, NADE and GCG in BI model from the $\Lambda$CDM, the $w$CDM and the  SGL+CBS models and  it was observed that the constraint on $\Omega_{\sigma 0}$ is very strong (see Figs.  (7) and (13)). For the case of  $\Omega_{\sigma 0}=0.02$,  $\triangle \nu$   was positive at small redshifts and  negative at  $z>0.64$, while for $\Omega_{\sigma 0}\neq0.02$, $\triangle \nu$   was negative in all range of redshift.
We have used the   Hubble parameter versus redshift data to constrain cosmological parameters of HDE  and NADE of GCG models in BI universe. The constraints are  consistent
with observational data than $\Lambda$CDM. In addition,  we show that  in anisotropic universe,  the HDE model is better than the NADE and GCG models  (see Fig.  (14)).
Also, Fig. (15) shows that the anisotropy would result in an evident influence on the cosmic evolution
by analyzing evolutionary  expansion rate $H(z)$. It was observed that the bigger   anisotropy is, the best value the
Hubble expansion rate $H(z)$ is gotten. Finally, we investigated the growth of structures in linear regime with effects of anisotropy and showed that the growth of density
perturbations $g(a)$ is slowed down in $\Lambda$CDM models compared to the HDE, NADE and GCG models (see Fig. (16)). So,
it is concluded that in an anisotropic universe the growth factor evolves more slowly with increasing the anisotropy
parameter and it will always fall behind the FRW universe.


\end{document}